\documentclass[aps,prd,eqsecnum,showpacs,amsmath,nofootinbib,twocolumn]{revtex4}

\usepackage{amsmath}
\usepackage[dvips]{color,graphicx}
\usepackage{amsfonts,amssymb,theorem,mathrsfs,times}

\newcommand{\D}{{\rm d}}
{\theorembodyfont{\upshape}

}
{\theorembodyfont{\upshape}

}
{\theorembodyfont{\upshape}

}
{\theorembodyfont{\upshape}

}
{\theorembodyfont{\upshape}

}
{\theorembodyfont{\upshape}

}
\newcommand{\dalm}{\kern1pt\vbox{\hrule height 0.9pt\hbox{\vrule width
0.9pt\hskip 2.5pt\vbox{\vskip 5.5pt}\hskip 3pt\vrule width 0.3pt}\hrule height
0.3pt}\kern1pt}

\begin{document}

\title{Can a black hole with conformal scalar hair rotate?}

\author{Sourav Bhattacharya$^{1}$}
\email{souravbhatta-at-hri.res.in}
\author{Hideki Maeda$^{2,3}$}
\email{hidekism-at-rikkyo.ac.jp}


\address{
$^{1}$Harish-Chandra Research Institute, Chhatnag Road, Jhunsi, Allahabad-211019, India\\
$^{2}$Department of Physics, Rikkyo University, Tokyo 171-8501, Japan\\
$^{3}$Centro de Estudios Cient\'{\i}ficos (CECs), Arturo Prat 514, Valdivia, Chile
}


\date{\today}

\begin{abstract}
It is shown that, under the separability assumption for the metric, the slow-rotation approximation for the Bocharova-Bronnikov-Melnikov-Bekenstein black hole in general relativity with a conformally coupled scalar field does not work outside the event horizon.
Suggestions indicated by our present analysis towards a fully rotating black hole solution are discussed.
\end{abstract}


\maketitle


\section{Introduction}
Conformally coupled scalar field has been paid much attention in black hole physics since an exact solution was found by Bocharova, Bronnikov, and Melnikov~\cite{Bocharova-Bronnikov-Melnikov} and independently by Bekenstein~\cite{Bekenstein1974} in the 70's.
It represents an asymptotically flat spherically symmetric black hole with a non-trivial configuration of the scalar field, namely a scalar hair.
This is quite intriguing because the black hole no-hair theorem has been proven for a wide class of scalar fields~\cite{Bekenstein:1996pn}.
In spite of the fact that the geometry is perfectly regular, the scalar field diverges on the event horizon, which does violate the key assumption of the no-hair theorem.

This so-called BBMB solution is the unique static solution with spherical symmetry~\cite{Klimcik1993}.
It is also true in arbitrary dimensions, but interestingly the resulting unique solution represents not a black hole but a naked curvature singularity in higher dimensions~\cite{Klimcik1993, Xanthopoulos:1992fm}. 
In this sense, the BBMB black hole is isolated in the dimensionality of spacetime.
The scalar-field singularity is indeed problematic to analyze the BBMB black hole.
Although it is harmless for particle motion even if it couples with the scalar field~\cite{Bekenstein1975}, it prevents us from performing the black hole thermodynamics~\cite{Zaslavskii2002} and also from finding a proper boundary condition on the horizon for perturbations. However, if the scalar hair is {\it a priori} assumed to be bounded on the horizon, there can be no asymptotically flat solution other than the Schwarzschild~\cite{Zannias:1994jf}.

Actually, this problem is resolved if we add a positive cosmological constant together with a quartic potential of the scalar field, which is required by the conformal coupling.
In this generalized solution, the scalar field singularity is hidden inside the event horizon and the temperature of the event and the cosmological horizons are equal~\cite{Martinez:2002ru}. 
We also refer our reader to~\cite{Galtsov:1992be} for an interesting solution-generating technique for this system.
Thermodynamics for this class of ``lukewarm" black holes can be performed in the Euclidean path integral approach~\cite{bdw2005}.

From this point of view, the geometry of the rotating BBMB black hole, if it exists, seems highly non-trivial and interesting.
Although the BBMB solution has been generalized for the case with a cosmological constant, quartic potential, Maxwell field, or different horizon topology~\cite{Virbhadra:1993st,Martinez:2002ru,Martinez:2005di}, the Kerr-like solution with scalar hair has not been obtained yet.
All of these spacetimes are included in the Pleba{\'n}ski-Demia{\'n}ski family, which is the most general Petrov type D spacetime in the Einstein-Maxwell system~\cite{Plebanski:1976gy,gp2006}.
It contains six parameters and represents an accelerating and rotating black hole in general.
Several years ago, the Pleba{\'n}ski-Demia{\'n}ski-type solution with scalar hair was obtained~\cite{am2010}, which contains five parameters and reduces to the BBMB black hole or the accelerating BBMB black hole~\cite{Charmousis:2009cm} in certain limits.
Quite recently, several efforts have also been made via solution-generating techniques~\cite{astorino2013b,astorino2013,bcc2013}.
However, the rotating BBMB black hole solution with scalar hair is still missing~\cite{am2010}.

In the present paper, we provide some suggestions for this problem by constructing the slowly-rotating BBMB solution.
Our basic notation is the following~\cite{wald}.
The convention for the Riemann curvature tensor is $[\nabla _\rho ,\nabla_\sigma]V^\mu ={R^\mu }_{\nu\rho\sigma}V^\nu$ and $R_{\mu \nu }={R^\rho }_{\mu \rho \nu }$.
The Minkowski metric is taken to be mostly plus sign, and Greek indices run over all spacetime indices.
We adopt the units in which only the gravitational constant $G$ is retained.

\section{Preliminaries}
\subsection{The system}
We consider general relativity with a cosmological constant $\Lambda$ and a conformally coupled scalar field, the action is given by 
\begin{align}
\label{action}
S=&\frac{1}{2\kappa}\int d^4x\sqrt{-g}(R-2\Lambda)+S_{\phi}, \\
S_{\phi}=&-\int \D^4x \sqrt{-{ g}} \left[\frac12({\nabla} \phi)^2 +\frac{1}{12}R\phi^2+\alpha \phi^{4}  \right],
\end{align}
where $\kappa:=8\pi G$ and $\alpha$ is constant.
This action gives the following field equations:
\begin{eqnarray}
G_{\mu \nu }+\Lambda g_{\mu \nu }&=&\kappa T_{\mu \nu }^{(\phi) },~~~~~~  \label{eqs} \\
T_{\mu \nu }^{(\phi) }&:=&(\nabla _{\mu }\phi)( \nabla_{\nu }\phi) -\frac{1}{2}g_{\mu \nu }(\nabla\phi)^2-\alpha g_{\mu \nu
}\phi ^{4} \nonumber \\
&&+\frac{1}{6}\left( g_{\mu \nu }\kern1pt%
\vbox{\hrule height 0.9pt\hbox{\vrule width
0.9pt\hskip 2.5pt\vbox{\vskip 5.5pt}\hskip 3pt\vrule width 0.3pt}\hrule height
0.3pt}\kern1pt -\nabla _{\mu }\nabla _{\nu }+G_{\mu \nu }\right) \phi ^{2},
\\
\kern1pt%
\vbox{\hrule height 0.9pt\hbox{\vrule width
0.9pt\hskip 2.5pt\vbox{\vskip 5.5pt}\hskip 3pt\vrule width 0.3pt}\hrule height
0.3pt}\kern1pt \phi&=&\frac16 R\phi+4\alpha \phi^3. \label{sc}
\end{eqnarray}

\subsection{BBMB black hole}
In the case of $\Lambda=\alpha=0$, the unique spherically symmetric static solution is the following BBMB solution~\cite{Bocharova-Bronnikov-Melnikov,Bekenstein1974}: 
\begin{align}
ds^2=&-f(r)dt^2+f(r)^{-1}dr^2+r^2d\Omega^2,\label{BBMB1}\\
f(r)=&\frac{(r-M)^2}{r^2},\qquad \phi=\pm\sqrt{\frac{6}{\kappa}}\frac{M}{r-M},\label{BBMB2}
\end{align}
where $M$ is a constant and $d\Omega^2:=d\theta^2+\sin^2\theta d\varphi^2$.
The metric is exactly the same as the extremal Reissner-Nordstr\"om spacetime and there is the scalar-field singularity on the event horizon $r=M$, where the spacetime is completely regular.

In the generalized solution~\cite{Martinez:2002ru} in the presence of $\Lambda$ and the quartic potential with $\alpha=-\kappa\Lambda/36$, the configuration of the scalar field remains the same but the metric function becomes 
\begin{align}
f(r)=\frac{(r-M)^2}{r^2}-\frac13\Lambda r^2.\label{BBMB2-lambda}
\end{align}
This spacetime contains an event horizon only for $\Lambda>0$ with
\begin{align}
0<M<\frac14\sqrt{\frac{3}{\Lambda}}. \label{horizon-M}
\end{align}
Under these inequalities, there are three non-degenerate Killing horizons, given by $f(r_{\rm h})=0$ and the mass-horizon relation is 
\begin{align}
M=r_{\rm h}\pm\sqrt{\frac{\Lambda}{3}}r_{\rm h}^2.
\end{align}
The plus sign is for the inner horizon while the minus sign is for the event horizon and the cosmological horizon.
The scalar field singularity at $r=M$ is located in the trapped region between the inner and event horizons.

\section{Slowly-rotating BBMB solution}

Let us consider the following slowly-rotating BBMB solution:
\begin{align}
ds^2=&-f(r)dt^2+f(r)^{-1}dr^2+r^2d\Omega^2 \nonumber \\
&~~~~~~~~~~~-2a \beta(r,\theta) dtd\varphi \label{BBMB1-slow2}
\end{align}
with the same $f(r)$ and $\phi(r)$ as in the BBMB black hole (\ref{BBMB2}).
Here $a~(a/M\ll 1)$ is the slow-rotation parameter.
In the slow-rotation approximation we adopt, we assume (i) $a$ is small and (ii) the metric function $\beta(r,\theta)$ is finite. 
We note here that we do not need to perturb the scalar field in the linear order in $a$, because stationarity and axisymmetry require that the rotation parameter $a$ in the scalar field appears with even power only, similar to the diagonal components of the metric.

\subsection{Separability}
The linearized field equations with $\alpha=-\kappa\Lambda/36$ give the following equation for $\beta(r,\theta)$:
\begin{align}
0=&3r^2(r-2M)(r-M)\biggl(\sin\theta \frac{\partial^2\beta}{\partial \theta^2}-\cos\theta \frac{\partial\beta}{\partial \theta}\biggl) \nonumber \\
&+\sin\theta\biggl[r^2(r-2M)(r-M)\biggl\{3(r-M)^2-r^4\Lambda\biggl\}\frac{\partial^2\beta}{\partial r^2} \nonumber \\
&+2M^2r\biggl\{3(r-M)^2-r^4\Lambda\biggl\}\frac{\partial \beta}{\partial r} \nonumber \\
&+2\biggl\{3M(r-M)(2r^2-7Mr+4M^2) \nonumber \\
&+\Lambda r^4(r^2-3Mr+4M^2)\biggl\}\beta\biggl],
\end{align}
which is a separable form.
Putting $\beta(r,\theta)=h(r)\Theta(\theta)$, we obtain
\begin{align}
0=&\sin\theta \frac{d^2\Theta}{d\theta^2}-\cos\theta \frac{d\Theta}{d\theta}+C\sin\theta\Theta , \label{beq1}\\
0=&r^2(r-2M)(r-M)\biggl\{3(r-M)^2-r^4\Lambda\biggl\}\frac{d^2h}{dr^2} \nonumber \\
&+2M^2r\biggl\{3(r-M)^2-r^4\Lambda\biggl\}\frac{dh}{dr} \nonumber \\
&+\biggl\{6M(r-M)(2r^2-7Mr+4M^2) \nonumber \\
&-3Cr^2(r-2M)(r-M)+2\Lambda r^4(r^2-3Mr+4M^2)\biggl\}h,\label{beq2}
\end{align}
where $C$ is the separation constant.
Defining $x:=\cos\theta$ of which domain is $-1\le x\le 1$, we rewrite the angular equation (\ref{beq1}) as
\begin{align}
0=&(1-x^2)\frac{d^2\Theta}{dx^2}+C\Theta.
\end{align}
The solution of this equation is given in terms of the Hypergeometric function ${}_2F_1(a_1,a_2,b,z)$ as
\begin{align}
\Theta(x) =& D_1(1-x^2) \nonumber \\
&\times {}_2F_1\biggl(\frac{3+\sqrt{1+4C}}{4},\frac{3-\sqrt{1+4C}}{4},\frac12,x^2\biggl) \nonumber \\
&+D_2x(1-x^2) \nonumber \\
&\times {}_2F_1\biggl(\frac{5-\sqrt{1+4C}}{4},\frac{5+\sqrt{1+4C}}{4},\frac32,x^2\biggl),
\end{align}
where $D_1$ and $D_2$ are constants.
Continuity of $\Theta(x)$ at $x=\pm 1$ requires $D_2=0$.
Then, the resulting $\Theta(x)$ is analytic at $x=\pm 1$ for any value of $C$.
The simplest case is with $C=2$, with which we obtain $\Theta =D_1(1-x^2)=D_1\sin^2\theta$.

\subsection{Regularity on the event horizon}
Let us first consider the case with $\Lambda=\alpha=0$.
The radial equation (\ref{beq2}) then becomes
\begin{align}
0=&r^2(r-2M)(r-M)^2\frac{d^2h}{dr^2}+2M^2r(r-M)\frac{dh}{dr} \nonumber \\
&+\biggl\{2M(2r^2-7Mr+4M^2)-Cr^2(r-2M)\biggl\}h.\label{beq2-2}
\end{align}
This equation is singular at $r=M$ and $r=2M$.
Suppose $h(r)$ is finite around $r=M$ and can be expanded as $h(r)\simeq h_0+h_1(r-M)^{p}$, where $p$ is a positive real number, we obtain, from the radial equation (\ref{beq2-2}), 
\begin{align}
0\simeq &(C-2)M^3h_0+(C-6)M^2h_0(r-M) \nonumber \\
&+(C-2)M^3h_1(r-M)^p-p(p-3)M^3h_1(r-M)^{p}. \label{expand1}
\end{align}
From the lowest order of the above equation, $C=2$ is concluded.
Then Eq.~(\ref{expand1}) reduces to 
\begin{align}
0\simeq &-4M^2h_0(r-M)-p(p-3)M^3h_1(r-M)^{p}
\end{align}
and hence $p=1$ and $h_1=2h_0/M$ are concluded.

This is also the case with positive $\Lambda$.
Equation (\ref{beq2}) is singular at $r=r_{\rm h}$ which is define by $3(r_{\rm h}-M)^2-r_{\rm h}^4\Lambda=0$.
Suppose $h(r)$ is finite around $r=r_{\rm h}$ and can be expanded as $h(r)\simeq {\bar h}_0+{\bar h}_1(r-r_{\rm h})^{q}$, where $q$ is a positive real number.
Then the radial equation (\ref{beq2}) gives
\begin{align}
0\simeq &-6q(q-1){\bar h}_1r_{\rm h}(r_{\rm h}-2M)^2(r_{\rm h}-M)^2(r-r_{\rm h})^{q-1} \nonumber \\
&-12q{\bar h}_1M^2(r_{\rm h}-2M)(r_{\rm h}-M)(r-r_{\rm h})^{q} \nonumber \\
&+\biggl\{3(2-C)r_{\rm h}^2(r_{\rm h}-2M)(r_{\rm h}-M)+\mathcal{O}((r-r_{\rm h})^1)\biggl\} \nonumber \\
&\times  \biggl({\bar h}_0+{\bar h}_1(r-r_{\rm h})^{q}\biggl)
\end{align}
around $r=r_{\rm h}$, which shows $q=1$ and $C=2$.

An alternative way to see this is the following. Since we are looking for a stationary black hole
spacetime, we must have a Killing horizon, where the function, say
$Z(r,\theta):=g_{t\varphi}/g_{\varphi\varphi}$ is constant so that the vector field
$\chi^\mu=(\partial_t)^\mu+Z(\partial_{\varphi})^\mu$ is Killing and null there (see 
eg.~\cite{Bhattacharya:2013caa} and references therein). Since the horizon is a $r={\rm constant}$
hypersurface, and $g_{\varphi\varphi}=r^2\sin^2\theta$, we must have $\beta(r,\theta)=\gamma h(r)\sin^2\theta$
uniquely everywhere, if we assume a variable separation,
where $\gamma$ is a constant. This constant can be absorbed in the 
rotation parameter `$a$' anyway, and the result follows.

\subsection{Non-existence}
In the previous subsection, we have shown that the finiteness of the metric function $h(r)$ at the event horizon requires $C=2$ and hence $\beta(r,\theta)=h(r)\sin^2\theta$.
Ignoring terms nonlinear in $a$ in the field equations, we obtain the governing equation (\ref{beq2-2}) for $h(r)$, which is solved to give
\begin{align}
h(r)=&c_1r^2+\frac{c_2}{r^2}\biggl[r^4\ln\biggl|1-\frac{2M}{r}\biggl| \nonumber \\
&~~~~~+2M(r-M)(r^2+2Mr-2M^2)\biggl] \label{sol-g}
\end{align}
in the case of $\Lambda=\alpha=0$, where $c_1$ and $c_2$ are constants.

Asymptotic flatness requires $c_1=0$.
The solution with $M>0$, $c_1=0$, and $c_2\ne 0$ represents the slowly rotating BBMB black hole.
However, this solution is not valid at or around $r=2M$, where $h(r)$ diverges and the assumption of slow-rotation is violated.

One might think of `pasting' the $c_1 r^2$ solution
in a neighborhood around $r=2M$, and then smoothly match it with the second solution for two
points at $r>2M$ and $r<2M$. Such matching must be done for the metric function and its first
and second derivatives to ensure the continuity of geodesics and curvature. However it is easy to
see by expanding the logarithm in the second solution for any $r>2M$ that such matching is
not possible.   

Actually, even in the presence of $\Lambda$ and quartic potential, the solution with linear $a$ is given by Eq.~(\ref{sol-g}).
In the case where Eq.~(\ref{horizon-M}) is satisfied with positive $\Lambda$, there are three horizons at $r=r_1,~ r_2,~r_{\rm c}~(r_1<r_2<r_{\rm c})$. The metric singularity at $r=2M$ in this case is located in the region $r_2<r<r_{\rm c}$, namely, in the untrapped region between the event horizon and the cosmological horizon.

Thus
we have seen that the slow-rotation approximation doest not work at $r=2M$ outside the event horizon.
It is interesting to note that there appears similar problem for a slowly rotating Boson star, too~\cite{Kobayashi:1994qi}.

\section{Conclusion}
 
In this paper, we have obtained the slowly rotating BBMB solution, under the separability assumption for the metric function $\beta(r,\theta)$ in Eq.~(\ref{BBMB1-slow2}).
What insights can we gain from this solution about the fully rotating black hole solution?

The metric of a general stationary and axisymmetric spacetime is written in the coordinates $(t,r,\theta,\varphi)$ as
\begin{align}
ds^2=&g_{tt}(r,\theta)dt^2+2g_{t\varphi}(r,\theta)dt d\varphi+g_{\varphi\varphi}(r,\theta) d\varphi^2 \nonumber \\
&+g_{rr}(r,\theta) dr^2+g_{\theta\theta}(r,\theta) d\theta^2, \label{eq:ansatz}
\end{align}
provided the Killing vectors generating stationarity and axisymmetry admit integral two-spaces orthogonal to the group orbits~\cite{exact}.
In the asymptotically flat case, we can identify the rotation parameter $a$ which is proportional to the ADM (or Komar) angular momentum.
The stationarity requires invariance for $a\to -a$ with $t\to -t$ and hence only $g_{t\varphi}$ contains $a$ with odd power in the form of $g_{t\varphi}=a{\bar g}_{t\varphi}$, while ${\bar g}_{t\varphi}$ and other metric functions contain $a$ with even power.

Actually, it is not difficult to construct a stationary and axisymmetric black hole solution in the present system.
The following BBMB solution with Taub-NUT charge is an example:
\begin{align}
ds^2=&-F(r)(dt+2n\cos\theta d\varphi)^2 \nonumber \\
&~~~~~~+F(r)^{-1}dr^2+(r^2+n^2)d\Omega^2,\\
F(r)=&\frac{(r-M)^2}{r^2+n^2},\\
\phi=&\pm\sqrt{\frac{6(n^2+M^2)}{\kappa}}\frac{1}{r-M},
\end{align}
where $n$ is the NUT parameter~\cite{comment}.
For positive $M$, this solution represents a black hole and reduces to the BBMB solution (\ref{BBMB1})--(\ref{BBMB2}) in the limit of $n\to 0$.
In this spacetime, $\lim_{r\to \infty}R^{\mu\nu}_{~~~\rho\sigma}= 0$ is satisfied but the fall-off conditions for asymptotic flatness~\cite{gravitation} is not respected.
Therefore, this spacetime is just asymptotically {\it locally} flat.

However, our chief interest is the asymptotically flat and fully rotating solution.
The problem in our present solution is the singularity in the metric function $h(r)$ at $r=2M$. 
Although it is not a curvature singularity at the linear level, it is still not clear whether the fully rotating solutions, if there are, contain a naked curvature singularity or not.
Also, there is still a possibility that rotating black hole solutions with non-separable $\beta(r,\theta)$ exist.
In order to shed light on the present problem, numerical analyses are quite useful, which will be reported elsewhere.


\subsection*{Acknowledgments}
The authors thank Tomohiro Harada for discussions and comments. 
SB thanks Rikkyo University, Tokyo, Japan, for its kind hospitality where initial discussions on this problem were done.  
This research was supported in part by the JSPS Grant-in-Aid for Scientific Research (A) (22244030). 
This work has been partially funded by the Fondecyt grants 1100328, 1100755 and by the Conicyt grant ``Southern Theoretical Physics Laboratory" ACT-91. 
The Centro de Estudios Cient\'{\i}ficos (CECs) is funded by the Chilean Government through the Centers of Excellence Base Financing Program of Conicyt.



{%

\end{document}